# Bury Me Here --The New Genre of Narrative Design Game Based on Immersive Storytelling

Zhongsheng Li[1]  Wuji Li[2]  and Yudong He[3]

[1] University of Texas at Austin, Austin TX 78705, USA
[2] University of California, Irvine
Irvine, CA, United States
[3] Huazhong University of Science and Technology
Wuhan, China

**Abstract.** Virtual reality games always provide the player with the most verisimilitude experience. With the advancement of VR hardware, it may become mainstream how people feel and attach to a virtual world. The paper discusses a possible solution to finding a better balance between the two classical genres of VR games, sensory stimulation and storytelling. To this end, we designed a game named "Bury Me Here," in which players can find an emotional bond between the game protagonist and themselves. The game includes four sections, the departure from the hometown, the travel on the train, the work in the office, and the life in the penthouse. At the game's end, the protagonist returns to his country yard and spends the rest of his life there. All the sections are designed to tell a stranger's life story to the player, making them experience someone else's life path and bonding an emotional connection between the player and the protagonist through storytelling. Results show that the game provides an immersive visual experience and has emotive sparks echo in players' minds.

## 1    Introduction

In a highly commercialized gaming industry nowadays, many big title games focus on flashy graphics and fast-paced gameplay. They are prone to marketing smokescreens and hype trains instead of thought-provoking emotional connections with the audience. Therefore, it is hard to associate those games with the idea of art.

According to a well-acclaimed film critic Roger Ebert, games will never be considered an art form because games have rules and goals, and they need to be won, and if a game does not have those, then it cannot be considered a game but rather a representation of a story [1]. The artist Yang Yongliang has an art piece titled Eternal Landscape that depicts an authentic Song-Yuan Mountain landscape. The art piece requires a Virtual Reality (VR) headset to function, and it was created through 3D modeling and 3D animations [2]. The audience can observe the landscape and even walk in it wearing the VR headset. This art piece technically utilizes the same game development methods, and it is considered not a game if we follow Ebert's ideology. However, by definition, video game refers to images being manipulated and interacted with through an electronic display. Furthermore, the fact that Eternal Landscape, an art piece that also technically counts as a video game [3], was exhibited in several different official art galleries is enough to disprove Ebert's claim that video games not being an art form.

In this work, titled "Bury Me Here," we wish to explore how games can be used as an impactful medium for storytelling and convey thoughtful emotions like art, film, photography, and many other similar mediums do. Since VR is being developed and implemented more and more thoroughly, we decided to put our project in VR as well. We intend to utilize environments, props, visuals, sounds, interaction, and reactions to provoke our participants' thoughts and emotions. The genre of our project would be described as a walking simulator, exploration, and visual novel. It is a familiar thing in the realm of video games. However, it fully utilizes the capabilities of VR, and the researchers wish to push the depth and length of emotional storytelling in VR as much as possible. The project also aims to work as a persuasion letter to people who think video games are less valuable than art and other similar floats.

## 2    Related Work

After the resurgence of VR devices, more and more VR products, games, and videos have been put on the stage, all of which are claimed to present new experiences to customers and audiences. Among all the user experiences mentioned, the most important is the immersive experience, which is also a keyword to catalog the previous work from the prior better [3]. An immersive experience in VR can be separated into two parts, immersive storytelling, and immersive sensory stimulation.

For immersive storytelling, a game called What Remains of Edith Finch shows how to use the first-person version to simulate the perspective of human walking. Walking simulation games often put 'narrative experience' before 'playing' [4]. Some people criticize this kind of work as being too monotonous and lacking in challenges at the interactive level [5]. It is hardly a game, more like a novel presented as a game. Also, in the game called The Vanishing of Ethan Carter, various superb cross-narrative techniques, foreshadowing, and metaphors were presented to the players, and various contradictions and doubts are intertwined, which also provides the possibility for further multiple interpretations and in-depth exploration. Besides the game, there is another immersive storytelling genre called immersive journalism [6]. A VR journal [7-8] produced by an emblematic group shows challenging social problem for their audiences, which includes domestic violence, police brutality, homophobia, and other political issues. Although the game is made in a low-poly scene, it still keeps its spirits. Like immersive storytelling, the genre lacks interaction between the project and the player and cannot take full advantage of the VR devices.

For immersive sensory stimulation, various games like the rollercoaster simulator and tree simulator take advantage of the parallax effect and 360 degrees sound effects provided by the VR devices, trying their best to provide the players with an immersive environment physically. Another fraction of the sensory stimulation is called visceral embodiment. In the VR video called The Extraordinary Honey Bee, users shrink to the size of a bee for a guided VR experience where they learn of the risks bee colonies face and solutions currently being implemented to offset their decline [9-10]. The virtual-reality project Tree transforms the player into a rainforest tree. With the players' arms as branches and the body as the trunk, the player could experience the tree's growth from a seedling into its fullest form and witness its fate firsthand [11-12]. The genre, like immersive sensory stimulation, focuses more on providing the player with a vivid and verisimilitude experience, short of being educational and meaningful at the same time. This trend makes most immersive sensory stimulation games lean toward violence, brutal subject, or simulation game just for fun. No matter which type of game it is, the key of those the writer mentioned above is offering an immersive and verisimilitude experience, creating empathy or mental stimulation.

## 3 Method

The researchers and designers used the Unity Engine to design the game and used Oculus quest2 to play the VR game. The designer focused on two effects to build the environment to bring the player a more immersive feeling.

### 3.1 Visual Effect

This project mainly applies visual effects in the farm scene. The techniques like Semi realism, Ambient Occlusion, Bloom, and Color Grading are implemented in modeling, rendering, and unique visual effect, which highly improves the game's immersion. For example, the plants and trees in the farm sway in the wind, and they will be surprised to find that the environment can be interacted with, which is just one of the details in this game but also consists of a harmonious environment. (**Figure 1**)

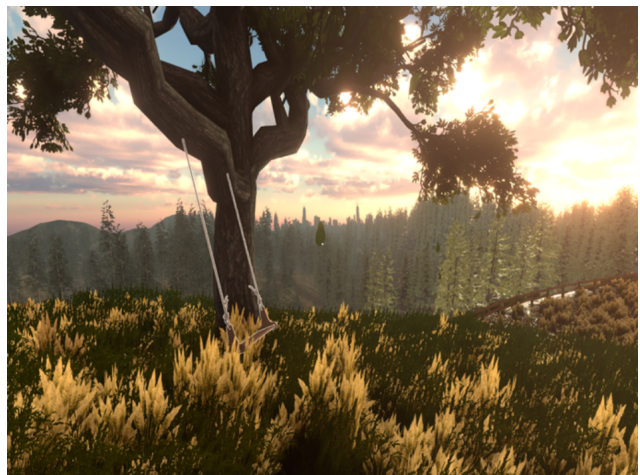

**Fig. 1.** Object interaction in the scene

In brief, such visual effects can directly convey the game's content to the players. In the visual narrative game, the visual effects will be paid more attention and play a more critical role in the immersion design.

### 3.2 Sound Effect

The sound effect is usually considered an auxiliary tool in game design. Although the sound is not applied to act as a primary way of gameplay in this game, it also highly contributes to the immersion design in some specific stages. In the farm scene, a guitar and corresponding sound effect is set in front of the hut, and the feature of the design is that the volume of the sound players can hear varies with the distance so that players will surprisedly feel like they are moving in reality.

## 4 Game Narrative Design

### 4.1 Overview

"Bury Me Here" is an immersive game about a period of ordinary but shining life. It would like to describe five stages of the protagonist's ordinary but shining life, from his childhood to maturity, who left the countryside for the big city for a better life. And finally, he found peace from the original humble life instead of wealth and fame, which is why he "bury me here" in the last stage.

This project is meant to connect with its audience on a deep emotional level. Therefore, the story needs to be as approachable as possible. Therefore, it would be best if the story's protagonist is just an ordinary person. The players visit/revisit the scenes in the protagonist's memories, and there are pieces of memories imbued in different objects inside each scene. (**Table 1**)

**Table 1.** The Plots with Emotion Stages

| Plots | Scenes | Emotion Stages |
|---|---|---|
| 1 | Farm | Eager to explore the world. |
| 2 | Carriage | Being on the road alone but still hopeful. |
| 3 | Office | Stuck in an endless loop of hell and can't see hope. |
| 4 | Penthouse | Wealth and fame but felt empty inside. |
| 5 | Farm | Going back to a humble life and finally found peace. |

It is up to the players to collect all pieces of memories and push along the narrative. When the players collect one memory, a background audio monologue will start to play as a way to tell the specific memory relating to the object. Since there is no explanation for who the protagonist is or whom the player is supposed to be, it is also up to the players to decide whether they are the protagonist or simply just an observer. The full story covers our protagonist's entire life, from childhood to death. So, with each scene forward, the player will follow the protagonist's story chronologically. However, other than the transition in time, it is also intended that players notice the different stages of the protagonist's emotions.

For the first stage on the farm, the protagonist would be in his childhood stage of yearning for adventures and explorations. As the plot goes on, the players would find themselves on a train carriage. This is the second stage which tells a story of the protagonist feeling lonely and adrift but still hopeful for a brighter future. Then the scene would transit into a more depressing environment, a confined office space, a cubicle of hell, representing the stage of our protagonist feeling stuck in life and hopeless. Nevertheless, life takes a better turn for our protagonist, who becomes wealthy and famous. The players would find themselves in the protagonist's luxurious penthouse and see how let-loose he has become and how this material life has made him emptier inside than ever. Finally, the journey ends where it begins, and our protagonist goes back to the old farm to find peace and the mundane in life, and that is where his story ultimately ends, just a normal guy who finally gets his humble ending.

### 4.2    Setting Details

#### Storyline

This game would like to describe five stages of the protagonist's ordinary but shining life, from his childhood to his maturity, who left the countryside for the big city for a better life. And finally, he found inner peace from the original humble life instead of wealth and fame.

The protagonist grew up on a plain farm in the countryside, where he spent an unforgettable childhood. With his gradual maturity, the player desired to explore the outside world. When he is on the train to the city, the anxiety about his uncertain future and his imagination of a better life intertwines in his mind. Fortunately, his diligence resulted in his success but also his indulgence in wealth and fame. He moved into a better house but was lost in this luxurious life. When he recalled his innocent experience in the countryside after he found his precious collection, he felt like coming back to where he bore in mind, and the story ended.

The project aims to connect with the audience on a deep emotional level, so the story needs to be as approachable as possible. Therefore, it is best to decide that the story's protagonist is just an ordinary person. The player visits and revisits scenes in the protagonist's memory, each containing a section of memory in different objects. Players need to collect all the memory fragments and drive the narrative. When the player collects a memory fragment, a background audio monologue will begin to inform the specific memory associated with the object. As each scene progresses, players will follow the protagonist's story chronologically. However, in addition to the transition in time, it is also intended to make the player notice the different stages of the protagonist's emotions.

**First Stage: Childhood Farm**
In this game's first scene, the protagonist returns to a farm in the countryside, where he carries his childhood memory with the family. This scene is supposed to be an introduction to the game, which can help players initially understand the basics and the following storyline. In the process of gameplay, players are guided to operate the protagonist to walk around to find a specific portal that transfers the protagonist to the next stage..

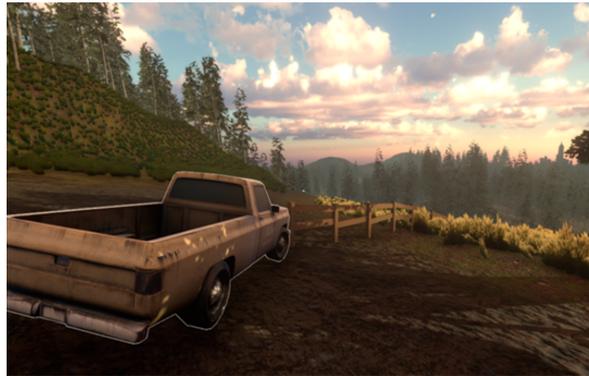

**Fig. 2.** Stage 1 Scenery Design

Involving the players in the game's narrative instead of making them spectators in the story can provide them with an immersive experience. To improve the immersion of the scene, most design methods, including visual effects, detail design, and interaction design, are appropriately implemented, accomplishing wonderful progress. For example, the plants and trees in the farm are set to sway in the wind, which is just one of the details in this game but also consists of a harmonious environment. **(Figure 2)** The complex design and foundation of the scenario creates the beautiful environment. In terms of object interaction, there is a swing interacting with players, which will play an exciting animation after players touch it. Figure 3 shows that a guitar and corresponding sound effects varying with the distance are set in front of the hut so that players can interact with it with VR equipment, experiencing the narrative more deeply. As a result, the emphasis researchers put on the environmental details significantly improves the perfection of players' experience.

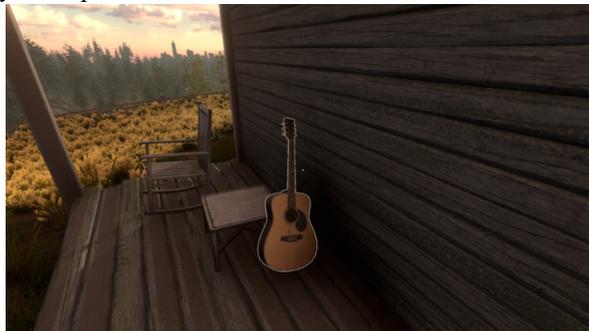

**Fig. 3.** Interaction Item to bond the feeling with players

**Second Stage: Confusing Train Journey**

The train scene is a transition between the farm towards the big city for our protagonist (**Figure 4**). He imagined his new life and threw all the troubles behind him on this train. The game quoted some sentences from 'On the Road,' like' What is that feeling when you're driving away from people, and they recede on the plain till you see their specks dispersing? — It is the too-huge world vaulting us, and it goodbye', hoping the narrative lines will help the player better understand the current feeling of the game protagonist.

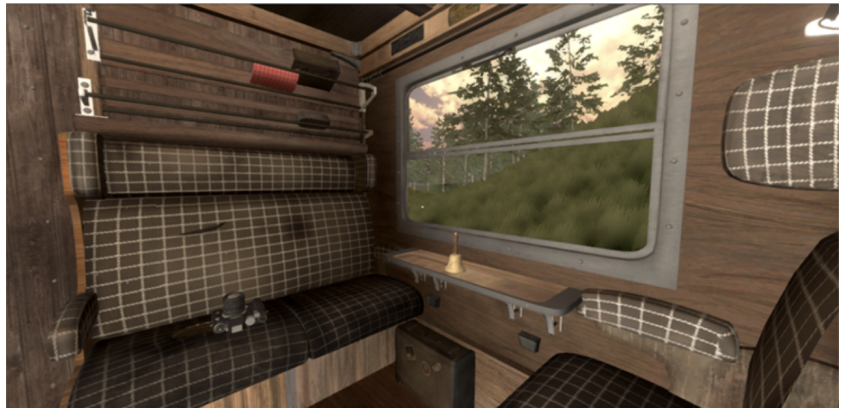

**Fig. 4.** Stage 2 Scenery Design

In the second stage, the train scene has abundant details in the carriage's interior, which shows the protagonist's experience and his mixed feelings when departing from the farm to the big city for a better life. As a result, the elaborate scene design can indicate the corresponding plot and background story, which improves players' immersion and provides them with a memorable gameplay experience with less work and space.

**Third Stage: Office**

In the third scene, the protagonist throws himself into his work with his American dream; he struggles in his small office，dreaming that one day he will return home with wealth and fame. However, the heavy workload turned him into a living working machine; all he cared about was the paperwork and numbers. He gradually forgot his hometown and wishes, working with a numb heart.

**Fourth Stage & Final Stage**

In the fourth scene, the protagonist comes to his luxurious penthouse, which indicates his success after hard work and struggle (**Figure 5**). But it also resulted in his indulgence in wealth and fame, which should be an important turn before the disenchantment of his life. Additionally, the obvious change in decorative style also symbolizes the transition of time and the new emotional state of the protagonist.

In the process of designing a game, echoing each other in front and behind is included to realize scene recall. In the fourth scene, the protagonist finds his old guitar, which brings back his memory of his innocent childhood. Then, this game returns to the previous farm that he bore in mind in these years, which echoes the game's theme. This structure would be vital feedback to the player's experience in the previous stages, which means they can be more immersive in presenting a stranger's life.

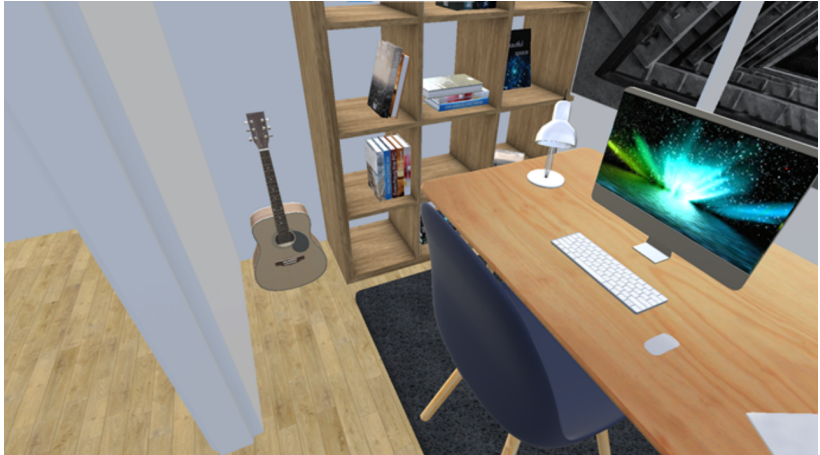

**Fig. 5.** Stage 4 Scenery Design

## 5      Conclusion & Discussion

This paper aims to discover how to apply visual narrative games to VR as an art form. We want to explore how games can be an impactful medium for storytelling and convey thoughtful emotions in the same way art, film, photography, and many other similar mediums do. The initial formulation of this project is to design an immersive game about a period of ordinary but shining life. It would like to describe five stages of the protagonist's ordinary but shining life, from his childhood to maturity, who left the countryside for the big city for a better life. And finally, he found peace from the original humble life instead of wealth and fame, which is why he "bury me here" in the last stage. The genre of our project would be described as a walking simulator, exploration, and visual novel. The elements of environments, props, visuals, sounds, interaction, and reactions are applied to stimulate the thoughts and emotions of our participants. This essay aims to fully utilize the capabilities of VR and push the depth and length of emotional storytelling as much as possible. Although the finished project lacks ample selective events or an open world, this paper specifies future development direction in the practical operation, exploring further developing space of this kind of game.

The link of the game demo video is  https://www.samuelleee.com/bury-me-here